\documentclass[12pt,preprint]{aastex}
 
\shorttitle{Detection of the HD189733b Exoplanet Atmosphere}
\shortauthors{Redfield et al.}

\begin{document}
 
\newcommand{\php}[0]{\phantom{--}}
\newcommand{\kms}[0]{km~s$^{-1}$}
 
\title{Sodium Absorption From the Exoplanetary Atmosphere of HD189733b
Detected in the Optical Transmission Spectrum\footnote{Based on
observations obtained with the Hobby-Eberly Telescope, which is a
joint project of the University of Texas at Austin, the Pennsylvania
State University, Stanford University,
Ludwig-Maximilians-Universit\"{a}t M\"{u}nchen, and
Georg-August-Universit\"{a}t G\"{o}ttingen.}}

\author{Seth Redfield\altaffilmark{2,3}, Michael Endl\altaffilmark{2}, William
D. Cochran\altaffilmark{2}, and Lars Koesterke\altaffilmark{2,4}}
\altaffiltext{2}{Department of Astronomy and McDonald Observatory,
University of Texas, Austin, TX, 78712; {\tt
sredfield@astro.as.utexas.edu}}
\altaffiltext{3}{Hubble Fellow}
\altaffiltext{4}{Texas Advanced Computing Center, University of Texas,
Austin, TX, 78758}

\begin{abstract}

We present the first ground-based detection of sodium absorption in
the transmission spectrum of an extrasolar planet.  Absorption due to
the atmosphere of the extrasolar planet HD189733b is detected in both
lines of the \ion{Na}{1} doublet.  High spectral resolution
observations were taken of eleven transits with the High
Resolution Spectrograph (HRS) on the 9.2 meter Hobby-Eberly Telescope
(HET).  The \ion{Na}{1} absorption in the transmission spectrum due to
HD189733b is $\left( -67.2 \pm 20.7 \right) \times 10^{-5}$ deeper in
the ``narrow'' spectral band that encompasses both lines relative to
adjacent bands.  The 1$\sigma$ error includes both random and
systematic errors, and the detection is $>$3$\sigma$.  This amount of
relative absorption in \ion{Na}{1} for HD189733b is $\sim$3$\times$ larger
than detected for HD209458b by \citet{charbonneau02}, and indicates
these two hot-Jupiters may have significantly different atmospheric
properties.

\end{abstract}
 

\section{Introduction}

High signal-to-noise ($S/N$) and high spectral resolution observations
of transiting exoplanets provide an opportunity to measure the
properties of exoplanet atmospheres and exospheres.  Early
theoretical models predicted that absorption, due to exoplanetary
atmospheres, should be detectable in the strong resonance lines of
\ion{Na}{1} and other alkali metals in optical transmission spectra
\citep{seager00,brown01}.  \citet{charbonneau02} used medium
resolution spectra ($R \equiv \lambda/\Delta\lambda \sim$ 5,540) of
\ion{Na}{1} from the Space Telescope Imaging Spectrograph (STIS)
onboard the {\it Hubble Space Telescope} ({\it HST}) to make the first
detection of absorption in the transmission spectrum caused by the
exoplanetary atmosphere of HD209458b during transit.  The measured
\ion{Na}{1} absorption, made in the ``narrow'' spectral window
(5887--5899\,\AA) since the doublet is not fully resolved, was
$\left(-23.2 \pm 5.7 \right) \times 10^{-5}$ deeper than in adjacent
bands during transit, and weaker than model predictions by a factor of
$\sim$3.  Adjustments to cloud levels, metallicity, rainout of
condensates, distribution of stellar flux, and photoionization of
sodium could account for the discrepancy between the models and
observations \citep{barman07,fortney03}.  Exospheric absorption was
also detected for HD209458b in several UV lines by
\citet{vidalmadjar03,vidalmadjar04}.

Previous ground-based attempts to detect absorption in the optical
transmission spectra of HD209458 and other bright transiting systems
have measured only upper limits, likely due to limited observations of
only a single effective transit
\citep[e.g.,][]{bundy00,moutou01,winn04,narita05,bozorgnia06}.  In
addition, contaminants, such as telluric lines and, in some cases, the
use of an iodine cell, hindered or prevented measurements of several
spectral features, most notably the \ion{Na}{1} lines.

%Radial velocity search spectra of
%HD209458, obtained by the High Resolution Echelle Spectrometer (HIRES)
%on Keck with an iodine cell, were used by \citet{bundy00} to place
%upper limits on calcium, hydrogen, and iron lines.  A partial transit
%of HD209458 observed with the Very Large Telescope (VLT) UV-Visual
%Echelle Spectrograph (UVES; $R\,\sim\,$58,000) was used by
%\citet{moutou01} to place limits on atmospheric and exospheric
%absorption.  HD209458 was observed with the High Dispersion
%Spectrograph (HDS) on Subaru ($R\,\sim\,$90,000) over a single transit
%and limits were placed on H$\alpha$ exospheric absorption
%\citep{winn04}, and absorption of \ion{Li}{1}, \ion{Fe}{1},
%\ion{Na}{1}, H$\beta$, H$\gamma$, \ion{Ca}{1} \citep{narita05}.
%Another bright transiting system, HD149026, was observed by
%\citet{bozorgnia06} with HIRES on Keck ($R\,\sim\,$55,000) over a
%single transit with an iodine cell, and limits were set on \ion{K}{1}
%and \ion{Li}{1} absorption.

\citet{bouchy05} announced the identification of the short-period
($P=2.2$ days) transiting exoplanet around the bright ($V = 7.7$) K0V
star, HD189733.  Subsequent work on this system tightly constrained
the orbital parameters \citep{hebrard06,bakos06}, determined that the
axis of stellar rotation is aligned with the planetary orbital axis
using the Rossiter-McLaughlin effect \citep{winn06}, found evidence
for photometric stellar variability and a stellar rotational period of
$\sim$12 days \citep{henry07,winn07}, and identified transit
signatures of starspots \citep{pont07}.  Infrared studies of this
system have been successful in characterizing the atmospheric
temperature distribution \citep{deming06,knutson07}.

\section{Observations and Data Reduction}

Based on the previous ground-based attempts to detect exoplanetary
absorption in transmission spectra, our program strategy was designed
to observe multiple transits and to specifically target
the strong \ion{Na}{1} lines.  Observations of HD189733 were taken
over the course of a year (11 Aug 2006 to 11 Aug 2007) using the
9.2~meter Hobby-Eberly Telescope (HET; \citealt{ramsey98}) High
Resolution Spectrograph (HRS; \citealt{tull98}).  These observations
cover the spectral range from 5000--9000~\AA, with a resolving power
of $R \sim$ 60,000.  No iodine cell was used.  A bright, rapidly
rotating, B star ($\alpha$ Del) was observed immediately after
HD189733 in order to characterize the telluric absorption.
We obtained spectra during 11 in-transit visits, and 25 out-of-transit
visits.  Each visit is typically comprised of several ($\sim$3.2
during transits and $\sim$5.1 outside of transits) 600 second
exposures, and the $S/N$ at \ion{Na}{1} of each exposure is
$\sim$320 per resolution element.  Observations are considered
in-transit, if the entire exposure is obtained within first and fourth
contact.  

%The fixed elevation-axis design of the HET necessitates
%queue-scheduling, which is well-suited to obtaining sporadic
%in-transit observations, together with the out-of-transit observations
%which can be acquired at the convenience of the telescope schedule.

The data were reduced using Image Reduction and Analysis Facility
\citep[IRAF;][]{tody93} and Interactive Data Language (IDL) routines
to subtract the bias, flat field the images, remove scattered light
and cosmic ray contamination, extract the echelle orders, calibrate
the wavelength solution, and convert to heliocentric velocities.
Local interstellar medium (LISM) and stellar lines were removed from
the telluric standard before being applied to the target spectra.
LISM \ion{Na}{1} absorption is detected in the spectra of $\alpha$ Del
($\sim$74 pc), our telluric standard, consistent with LISM detections
toward other nearby stars \citep{redfield02} and previous \ion{Ca}{2}
and \ion{Na}{1} observations of $\alpha$ Del by \citet{vallerga93} and
\citet{welsh91}.  Stellar lines in our telluric standard, a rapidly
rotating hot B-star, are very weak, and were removed using a stellar
model \citep{koesterke07,fitzpatrick05,castelli06}.  Observations of
HD189733 were divided by the cleaned telluric standard spectra to
remove the contaminating telluric features.

To create the in-transit and out-of-transit templates, additional
processing was required to match the blaze response and wavelength
calibration of all observations.  Over the short spectral windows
($\sim$30 \AA) of interest, fluctuations in the blaze response are
smooth and slowly varying, and were easily characterized by a low
order polynomial.  Likewise, precise spectral alignment required small
adjustments ($\sim$0.1--0.2 pixels) to the wavelength solution based
on the cross correlation of nearby stellar lines.

\section{\ion{Na}{1} Transmission Absorption}

Figure~\ref{fig:na} shows the observed \ion{Na}{1} region of the
stellar spectrum.  The doublet is fully resolved and accompanied by
many weak stellar lines.  The bottom plot shows the transmission
spectrum, or the difference of the relative fluxes, $\left({\cal
F}_{\rm in} - {\cal F}_{\rm out}\right)/{\cal F}_{\rm out}$, of the
ultra high $S/N$ in-transit template (${\cal F}_{\rm in}$; $S/N \sim
1600$) and the even higher $S/N$ out-of-transit template (${\cal
F}_{\rm out}$; $S/N \sim 3400$).  The signal predicted due to
differential stellar limb-darkening is shown, and has been removed
from the observations, although the impact is minimal.

Excess absorption is clearly detected in the in-transit spectrum
(${\cal F}_{\rm in}$) compared to the out-of-transit spectrum (${\cal
F}_{\rm out}$) for both \ion{Na}{1} lines.  Using the same ``narrow''
spectral region defined by \citet{charbonneau02}, 5887--5899~\AA, we
measure an excess of absorption $\left( -67.2 \pm 7.2 \right) \times
10^{-5}$, when compared to the adjacent spectrum.  This is $\sim$3$\times$
larger than measured for HD209458b \citep[{[$-23.2 \pm 5.7] \times
10^{-5}$;}][]{charbonneau02}.  We measure a random error of $7.2 \times
10^{-5}$, comparable to that achieved by the STIS/{\it HST}
observations.

We also analyze a neighboring strong stellar line that is predicted to
show no significant exoplanetary atmospheric absorption.  In
Figure~\ref{fig:ca}, we present the same analysis near the \ion{Ca}{1}
line at 6122\,\AA.  No absorption is detected in this
region, and over a narrow spectral range from 6119--6125\,\AA, we
measure an insignificant amount of emission $\left( +15.4 \pm 6.8
\right) \times 10^{-5}$, consistent with there being no difference
between the in-transit and out-of-transit templates.

\subsection{Stellar Limb Darkening}

Differential limb-darkening, the relative difference in the
limb-darkening response of the cores of spectral lines and the
continuum, contributes to the transmission spectrum.  Since we are
dealing with a narrow spectral window, and the spectra are normalized
by the adjacent continuum, these observations are not sensitive to the
broadband color dependency expected due to limb-darkening
\citep{brown01etal,ballester07}.

Essentially, the in-transit template is a collection of spectra where
a small fraction, ($\sim$2.5\%; \citealt{pont07}), of the stellar disk
is blocked.  This blocked component causes a loss of flux at a
particular radial velocity, and leads to the Rossiter-McLaughlin
effect \citep{winn06}.  In addition, the blocked component is
associated with a specific $\mu$-angle, and therefore causes the loss
of a specifically limb-darkened fraction of the stellar disk.  The
observational distribution of $\mu$ angles, all necessarily $<$0.73
since the HD189733b transits at an angle slightly inclined from
precisely edge-on \citep[$i=85.7^{\circ}$;][]{pont07}, will produce a
unique differential limb-darkening signature.  Indeed, with sufficient
$S/N$, a transiting exoplanet could be used as a blocking obstruction,
where the resulting differential limb-darkening spectrum would
constrain, among other things, the temperature profile of the stellar
atmosphere.  In our case, we are concerned about the contribution this
signal may have on the absorption feature in the transmission
spectrum.

We use a stellar model of HD189733, based on literature values,
$T_{\rm eff} = 5050$ K, $\log g = 4.5$, and Fe/H$ = -0.03$
\citep{bouchy05}, to calculate the stellar flux blocked by the
exoplanet for each observation \citep{koesterke07}.  We assume a 0.755
$R_{\odot}$ star and a 1.154 $R_{\rm Jup}$ planet \citep{pont07}.  The
blocked flux is subtracted from the integrated disk flux and the
continuum is normalized.  The colored spectrum in the bottom panel of
the left plot in Figures~\ref{fig:na} and \ref{fig:ca}, shows the
relative difference spectrum due to differential limb-darkening for
our distribution of in-transit observations.  The contribution of
differential stellar limb-darkening is much smaller than the observed
absorption, and as for HD209458 \citep{charbonneau02}, at \ion{Na}{1},
it is actually a net emission, where it contributes $+1.46 \times
10^{-5}$ in the ``narrow'' spectral window.  At \ion{Ca}{1} the
contribution is $-0.10 \times 10^{-5}$.  In both cases, the
differential limb-darkening contribution is well below the 1$\sigma$
random error.

\subsection{Empirical Monte Carlo \label{sec:emc}}

Given the weakness of any atmospheric absorption in a transmission
spectrum, it is critical to assess the contribution of systematic
errors.  Systematic errors associated with data reduction may arise
from the continuum normalization, wavelength calibration, 
removal of the stellar and interstellar lines from the telluric
standard, and the differential limb-darkening calculation.  In
addition, there are systematic contributions that are unrelated to
data reduction procedures, including stellar variability, the changing
distribution of starspots, and transits over active regions.  These
issues are particularly important since HD189733 is a moderately
active star that is known to exhibit photometric variability due to
star spots \citep{henry07,pont07}.  Furthermore, because our dataset
is comprised of many observations, spread over a year, it is important
to quantify the stability of the individual observations that
contribute to our in-transit and out-of-transit templates.

We have constructed a diagnostic that we call an empirical Monte
Carlo, or bootstrapping, analysis.  It involves randomly selecting
from the $\sim$200 individual exposures a test sample of
``in-transit'' exposures, running through our entire data analysis
algorithm, and measuring any signal in the transmission spectrum.  We
repeat this exercise thousands of times in order to establish
statistical significance.  Three different scenarios are explored: (1)
an ``out-out'' comparison, where a subset of out-of-transit exposures,
equal to the number of our in-transit exposures ($\sim$36), is
selected and compared to the rest of the out-of-transit exposures; (2)
an ``in-in'' comparison, where a subset of in-transit exposures
($\sim$8), set at the same ratio as the number of in-transit exposures
to the number of out-of-transit exposures (36/165), is compared to the
rest of the in-transit exposures; and (3) an ``in-out'' comparison,
where an increasing number of in-transit exposures (up to $\sim$half
of the sample) is removed from the in-transit template and compared to
the out-of-transit template.  The distribution of the ``narrow'' band
difference of the relative flux measurements for all realizations of
the three scenarios are shown for \ion{Na}{1} in the right panel of
Figure~\ref{fig:na} and for \ion{Ca}{1} in Figure~\ref{fig:ca}.

Both the ``in-in'' and ``out-out'' distributions are centered on zero
net signal, indicating that there is no significant difference between
the different out-of-transit exposures (``out-out'' scenario), and no
significant difference between the various in-transit exposures
(``in-in'' scenario).  Since we have fewer in-transit exposures in
order to do an ``in-in'' comparison, the $S/N$ is greatly reduced, and
the distribution is much wider than the ``out-out'' distribution.  For
\ion{Na}{1}, the mean value ($\mu$) of the ``out-out'' distribution is
$-3.5 \times 10^{-5}$ with a standard deviation ($\sigma$) of $20.7
\times 10^{-5}$, whereas the ``in-in'' distribution $\mu = 3.9 \times
10^{-5}$ and $\sigma = 35.5 \times 10^{-5}$.  The results are similar
for the \ion{Ca}{1} line where for the ``out-out'' distribution $\mu =
-0.5 \times 10^{-5}$ and $\sigma = 13.7 \times 10^{-5}$, and the
``in-in'' distribution $\mu = -4.0 \times 10^{-5}$ and $\sigma = 32.8
\times 10^{-5}$.  The \ion{Ca}{1} line is located more centrally on
the blaze and therefore is acquired with slightly higher $S/N$, which
translates into a narrower distribution of measurements.

As expected, the ``in-out'' distributions are centered on the value we
obtained from the full dataset.  The width of this distribution
provides a consistency check that a handful of in-transit exposures
are not dominating the measurement.  This demonstrates, along with the
``in-in'' distribution being centered at zero, that all the in-transit
exposures, over 11 transits, consistently show excess absorption from
the transiting exoplanetary atmosphere.

The width of the ``out-out'' distribution measures directly the
propagation of systematic errors through our analysis.  These
realizations include the contribution of both reduction systematics,
as well as any astrophysical systematics which are not correlated with
the exoplanet orbital period.  For example, if changing distributions
of starspots (i.e., different phases of the stellar variability cycle)
could mimic an absorption feature in the transmission spectrum, we
would expect to see a significant fraction of ``out-out'' realizations
result in such measurements.  Because no such realizations are able to
replicate the level of the detection, and because the orbital period
is completely decoupled from the stellar variability period
\citep{henry07} such that our 11 in-transit visits are randomly
distributed in stellar rotational phase, stellar variability cannot be
responsible for the observed absorption in the transmission spectrum.
The standard deviation of our ``out-out'' distribution is adopted as
the appropriate 1$\sigma$ error for our transmission spectrum
measurements since it encompasses both random and systematic errors.
Our \ion{Na}{1} absorption detection is $\left( -67.2 \pm 20.7 \right)
\times 10^{-5}$, while the signal at \ion{Ca}{1} is $\left( +15.4 \pm
13.7 \right) \times 10^{-5}$.

\section{Properties of the HD189733b Atmosphere}

\begin{figure}
\epsscale{1.2}
\plottwo{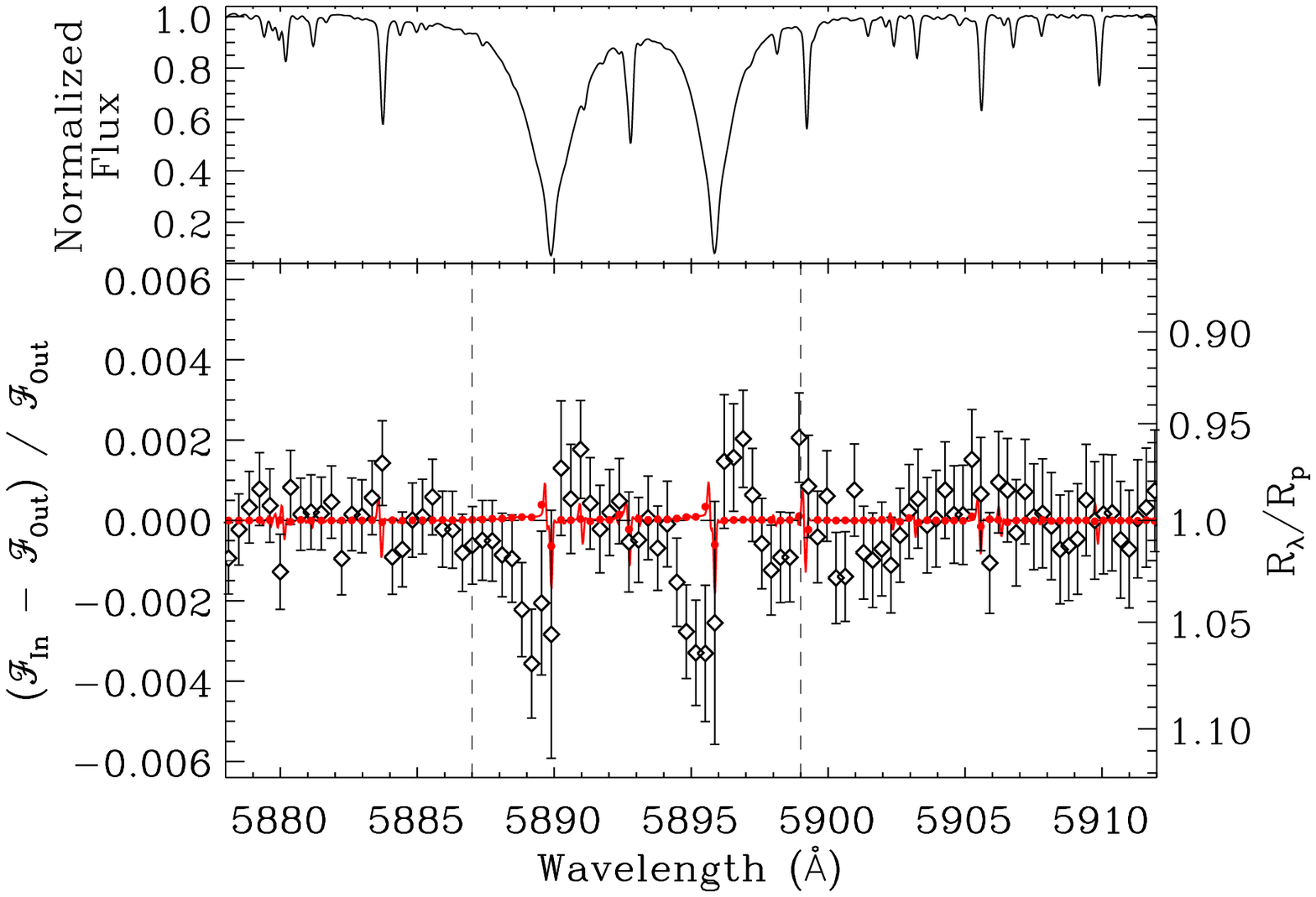}{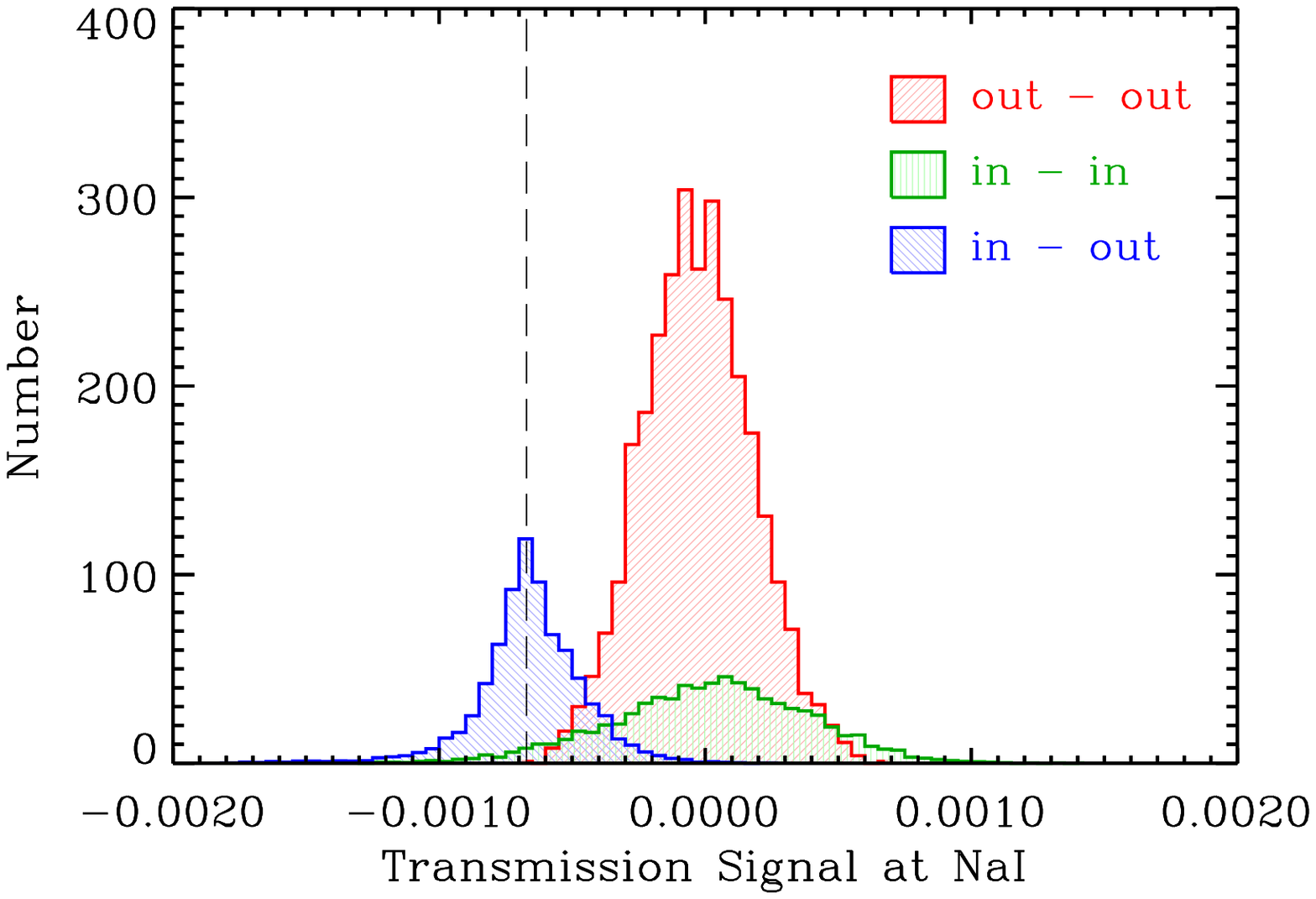}
\caption{{\it top left:} Spectrum of HD189733 near the \ion{Na}{1}
doublet.  {\it bottom left:} The difference of the relative flux of
the in-transit template (${\cal F}_{\rm in}$) and the out-of-transit
template (${\cal F}_{\rm out}$).  Excess absorption in the in-transit
template is clearly detected in both \ion{Na}{1} lines.  Also shown is
the contribution of differential limb darkening (red line and filled
circles), which has been removed from the data, but the impact is
minimal.  The vertical dashed lines indicate the narrow bandpass used
to calculate the relative transmission spectrum absorption, and is
identical to that used by \citet{charbonneau02}.  The righthand axis
gives the effective radius of the planet as a function of wavelength,
assuming stellar and planetary properties from \citet{pont07}.  {\it
right:} Distributions of the empirical Monte Carlo analysis, which
demonstrate the stability of the observations and the contribution of
systematic errors.  The full dataset transmission spectrum absorption
measurement is indicated by the vertical dashed line.  Given the
random and systematic errors measured by the standard deviation of the
``out-out'' distribution, this exoplanetary absorption detection is
$>$3$\sigma$.\label{fig:na}}
\end{figure}

\begin{figure}
\epsscale{1.2}
\plottwo{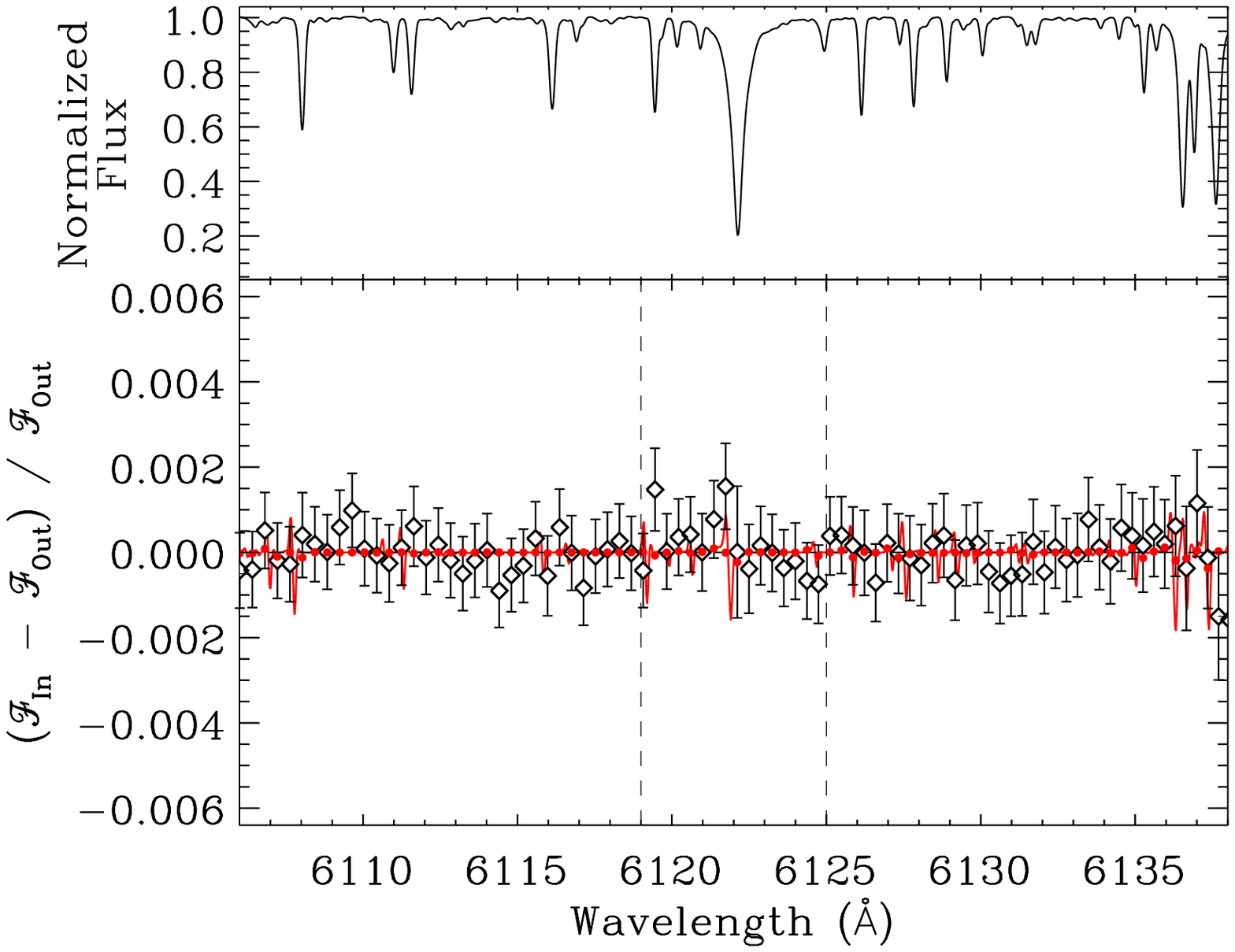}{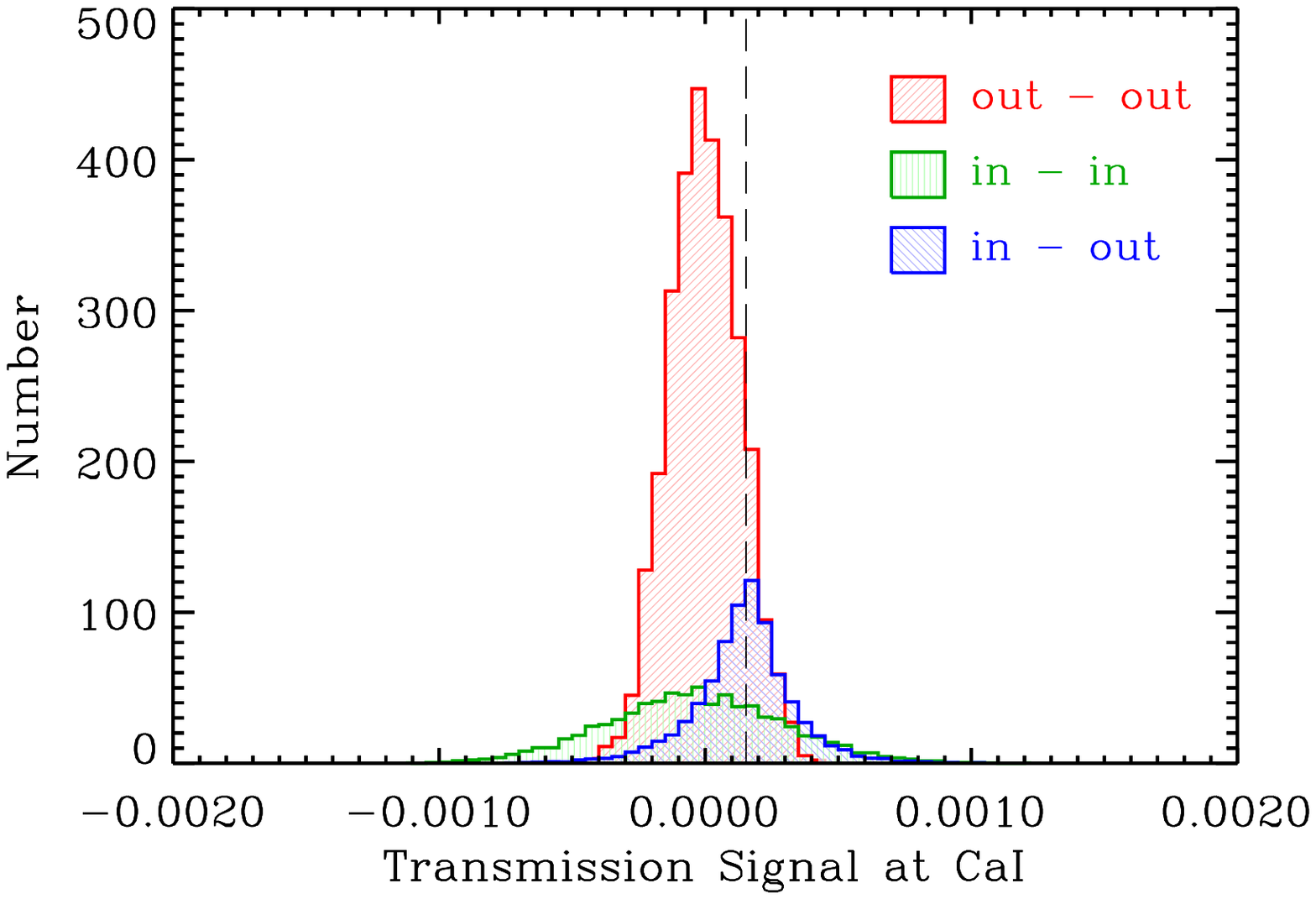}
\caption{Same as Figure~\ref{fig:na}, but for the \ion{Ca}{1} line,
which like \ion{Na}{1} is a strong line, but predicted not to show
absorption in the transmission spectrum.\label{fig:ca}}
\end{figure}

The absorption feature in the transmission spectrum in
Figure~\ref{fig:na} is a direct measurement of the effective radius of
the exoplanetary atmosphere due to the high opacity at the center of
the strong \ion{Na}{1} lines.  The right axis of Figure~\ref{fig:na}
shows the magnitude of the absorption in units of exoplanetary radius.
The cores of the \ion{Na}{1} lines extend to effective radii of
$\sim$1.06~$R_{\rm pl}$.

The spectra have not been corrected for the Doppler shift of the
exoplanetary absorption.  During conjunction, (e.g., mid-transit), the
radial velocity of the planet is 0 km\,s$^{-1}$, although the rate of
change in radial velocity is at its maximum.  Over the duration of the
transit, the radial velocity of HD189733b will change by $\pm$19.2
km\,s$^{-1}$.  This will cause a smearing of the absorption profile.
The observed feature appears to be significantly blueshifted from the
stellar line center, by $\sim$38 km~s$^{-1}$, and may be a combination of the planetary orbital motion and high speed winds flowing from the hot dayside, predicted by several dynamical atmospheric models \citep[e.g.,][]{cho03,cooper05}.

%The blueshift is
%apparent in all transmission spectra created from subsets of the full
%in-transit dataset (all realizations that make up the ``in-out''
%distribution in Figure~\ref{fig:na}), and even subsets binned as a
%function of transit phase.

Recent infrared observations of HD209458b by \citet{knutson07inver}
show evidence for a temperature inversion in the upper atmosphere of
that exoplanet.  \citet{burrows07} presented accompanying models which
require an additional absorber in the upper atmosphere of HD209458b,
which could be high-altitude clouds and explain the weaker than
expected \ion{Na}{1} absorption signal in the transmission spectrum
\citep{charbonneau02,fortney03}.  Our measurement of strong
\ion{Na}{1} absorption argues against the presence of such clouds high
in the atmosphere of HD189733b, consistent with the isothermal
temperature profile presented for HD189733b \citep{tinetti07}.

\section{Conclusions}

We present the first ground-based detection of absorption in the
transmission spectrum of a transiting extrasolar planet.  The
detection is based on the first high resolution ($R \sim 60$,000)
transmission spectrum, with the \ion{Na}{1} doublet fully resolved.
The \ion{Na}{1} transmission spectrum absorption for HD189733b,
measured in a ``narrow'' spectral band that encompasses both lines, is
$\left( -67.2 \pm 20.7 \right) \times 10^{-5}$.  The estimated error
includes both random and systematic contributions.  The amount of
absorption in \ion{Na}{1} due to HD189733b is $\sim$3$\times$ larger
than detected for HD209458b by \citet{charbonneau02}, and indicates
that the two exoplanets may have significantly different atmospheric
properties.  The excess absorption appears in a blue shifted
component, while the red half of the Na I doublet lines has a net
deficit of absorption during transit.  Future work will include the
analysis of other important optical lines (e.g., \ion{K}{1},
H$\alpha$), detailed analysis of the absorption profile and comparison
with atmospheric models, measurements of spectral indicators of
stellar variability, and observations of other transiting systems.
High resolution space-based spectroscopic observations will provide a
reliable confirmation of this detection, and access to important lines
in the ultraviolet.  Transmission spectra do not only have the
potential to probe the physical properties of hot-Jupiter and
hot-Neptune exoplanets, but lay the foundation toward detailed
physical diagnostics of Earth-sized extrasolar planetary atmospheres
\citep{ehrenreich06}.

\acknowledgements 

The authors would like to thank the referee, Ronald Gilliland, for his
insightful comments.  S.R. would like to acknowledge
support provided by the Hubble Fellowship grant HST-HF-01190.01
awarded by the STScI, which is operated by the AURA, Inc., for NASA,
under contract NAS 5-26555.  M.E. and W.D.C. acknowledge support
provided by NASA under grants NN05G107G and NNX07AL70G issued through
the TPF Foundation Science and the Origins of Solar Systems
programs. The Hobby-Eberly Telescope (HET) is a joint project of the
University of Texas at Austin, the Pennsylvania State University,
Stanford University, Ludwig-Maximilians-Universit\"{a}t M\"{u}nchen,
and Georg-August-Universit\"{a}t G\"{o}ttingen. The HET is named in
honor of its principal benefactors, William P. Hobby and Robert
E. Eberly.

{\it Facilities:} \facility{HET (HRS)}

%figures
\clearpage

%figures
\clearpage

\end{document}